\documentclass[aps,prl,twocolumn,superscriptaddress]{revtex4}
\usepackage{graphicx}
\usepackage{amsmath}
\usepackage{amssymb}
\usepackage{pifont}
\usepackage[dvipsnames]{color}
\usepackage{gensymb}
\usepackage{subeqnarray}

\begin{document}

\title[Activitiy Coefficients]{Simple Theory of Ionic Activity in Concentrated Electrolytes}

\author{Sven Schlumpberger}
\affiliation{Department of Chemical Engineering, Massachusetts Institute of Technology, Cambridge, MA 02139, USA}

\author{Martin Z. Bazant}
\affiliation{Department of Chemical Engineering, Massachusetts Institute of Technology, Cambridge, MA 02139, USA}
\affiliation{Department of Mathematics, Massachusetts Institute of Technology, Cambridge, MA 02139, USA}
\email{bazant@mit.edu}

\date{\today}


\begin{abstract}
The Debye-H\" uckel formula for ionic activity coefficients is extended for concentrated solutions by solving a simple model of many-body Coulomb correlations and adding the Born solvation energy.   Given the bulk permittivity, our formula is able to fit activity data for diverse electrolytes with only one parameter to adjust the correlation length, which interpolates between the Bjerrum length and the mean ion spacing. The results show that ionic activity in most electrolytes is dominated by three types of electrostatic forces: (i) mean-field charge screening, (ii) solvation, and (iii) Coulomb correlations, both ``over-screening" (charge oscillations) and ``under-screening" (extending beyond the Debye screening length).
\end{abstract}

\maketitle




The theory of ionic activity has a long history \cite{lewis_activity_1921,onsager_theories_1933,kirkwood_theory_1934,
scatchard_concentrated_1936,stokes_ionic_1948,pitzer1977electrolyte,newman_kirkwoodbuff_1989} since the seminal paper of Debye and H\"uckel (DH) in 1923~\cite{debye_theorie_1923}, but a simple physical model for concentrated solutions remains elusive.  Thermodynamic calculations~\cite{may2017thermodynamic,zemaitis2010handbook,tester_thermodynamics_1997,bromley_thermodynamic_1973} and engineering models~\cite{newman_electrochemical_2012} are usually based on empirical  formulae, such as the Pitzer equation with fitted second virial coefficients \cite{pitzer1973thermodynamics,kim1988evaluation,clegg1992thermodynamics}.  On the other hand, statistical theories are too complicated to solve analytically, even for charged hard spheres \cite{hansen_book,sloth1990single,molero1992individual,outhwaite1993primitive,simonin1998real,caccamo1996integral}, and still contain adjustable parameters.

The DH formula successfully predicts the activity coefficient $\gamma_i$ or excess chemical potential, $\mu_i^{ex}=k_BT \ln \gamma_i$, of species $i$ in a dilute solution via a mean-field approximation of the Coulomb energy between each ion and its correlated ``screening cloud" of excess counter-charge~\cite{debye_theorie_1923},
\begin{equation}
\ln \gamma_i^{DH} = -\frac{\left(z_ie\right)^2}{8\pi\varepsilon k_BT \left(a_i+\lambda_D\right)}   \label{eq:DH}
\end{equation}
in terms of the effective ionic radius $a_i$ (a fitting parameter), the ionic charge $z_ie$, the solution permittivity $\varepsilon$, and the Debye screening length, $\lambda_D = \sqrt{\varepsilon k_BT/2e^2I}$, where $I = \frac{1}{2}\sum_i z_i^2 s_i c_0$ is the ionic strength, 
$k_BT$ the thermal energy, $c_0$ the bulk salt concentration, and $\bar{c}_i=s_i c_0$ the bulk concentration of species $i$. The DH activity coefficient decreases with increasing salt  concentration, as attractive Coulomb correlations become stronger, but breaks down for $c_0>0.1$M and fails to predict the enhanced activity of most (but not all) electrolytes at high concentrations.  

In 1925, H\"uckel \cite{huckel1925theorie} proposed adding the change in self-energy of solvation to the DH formula,
\begin{equation}
\ln\gamma_i^H = \ln \gamma_i^{DH} + \frac{\Delta E_i^B}{k_BT}
\label{eq:DHB}
\end{equation}
in the approximation just introduced by Born~\cite{born_volumen_1920}
\begin{equation}
\Delta E_i^B = \frac{(z_ie)^2}{8\pi a_i}\left(\frac{1}{\varepsilon}-\frac{1}{\varepsilon_s}\right)
\label{eq:Born}
\end{equation}
which is the change in electrostatic potential of an isolated charged sphere, as the permittivity varies from $\varepsilon_s$ for pure solvent ($=78.36 \varepsilon_0$ for water) to $\varepsilon$ for the solution.  Since the local electric field of each ion aligns nearby solvent dipoles, the solution permittivity decreases with increasing concentration, thereby hindering solvation and increasing the activity.  H\"uckel linearized the Born energy for small dielectric decrements, $\Delta E_i^B \sim k_i I$, and fitted early activity data~\cite{huckel1925theorie}.  Although  H\"uckel's theory was later found to be consistent with measured permittivities~\cite{hasted1948dielectric}, the ``extended DH equation" with a linear term in ionic strength~\cite{meier1982two,rowland2014thermodynamics} and the ``specific ion interaction theory" (SIT) with linear terms for each ion concentration~\cite{may2017thermodynamic},
are widely viewed as empirical ~\cite{wright_book}, having lost their original connection with solvation energy, Eq. (\ref{eq:DHB}).

Since the 1960s, theorists shifted their attention from solvation energy to Coulomb correlations and excluded volume, using molecular simulations and more complicated statistical theories. Many calculations based on the Primitive Model~\cite{friedman1960mayer,hansen_book} (charged hard spheres in an implicit solvent of constant permittivity)  
predicted activity coefficients that were larger than DH and sometimes increasing~\cite{molero1992individual,outhwaite1993primitive} but typically decreasing ~\cite{simonin1998real,caccamo1996integral,sloth1990single} with ionic strength.  The field has recently come full circle, as Vincze et al.~\cite{vincze_nonmonotonic_2010} discovered that adding the Born energy could reverse this trend and bring hard-sphere simulations in line with experimental data.  More sophisticated ``molecular DH theories" including dielectric response have since emerged~\cite{xiao2011molecular,xiao2015analytical}, but still no formula of comparable simplicity and ease of interpretation as DH, with the notable exception of Fraenkel's ``smaller ion shell" (SIS) extension of DH~\cite{fraenkel2010simplified,fraenkel2011comment,vincze2011response,fraenkel2012single,fraenkel2014computing}, which approximates packing constraints in the screening cloud for size-dissimilar ions at moderate dilution.

\begin{figure*}[ht]
\includegraphics[width=7in]{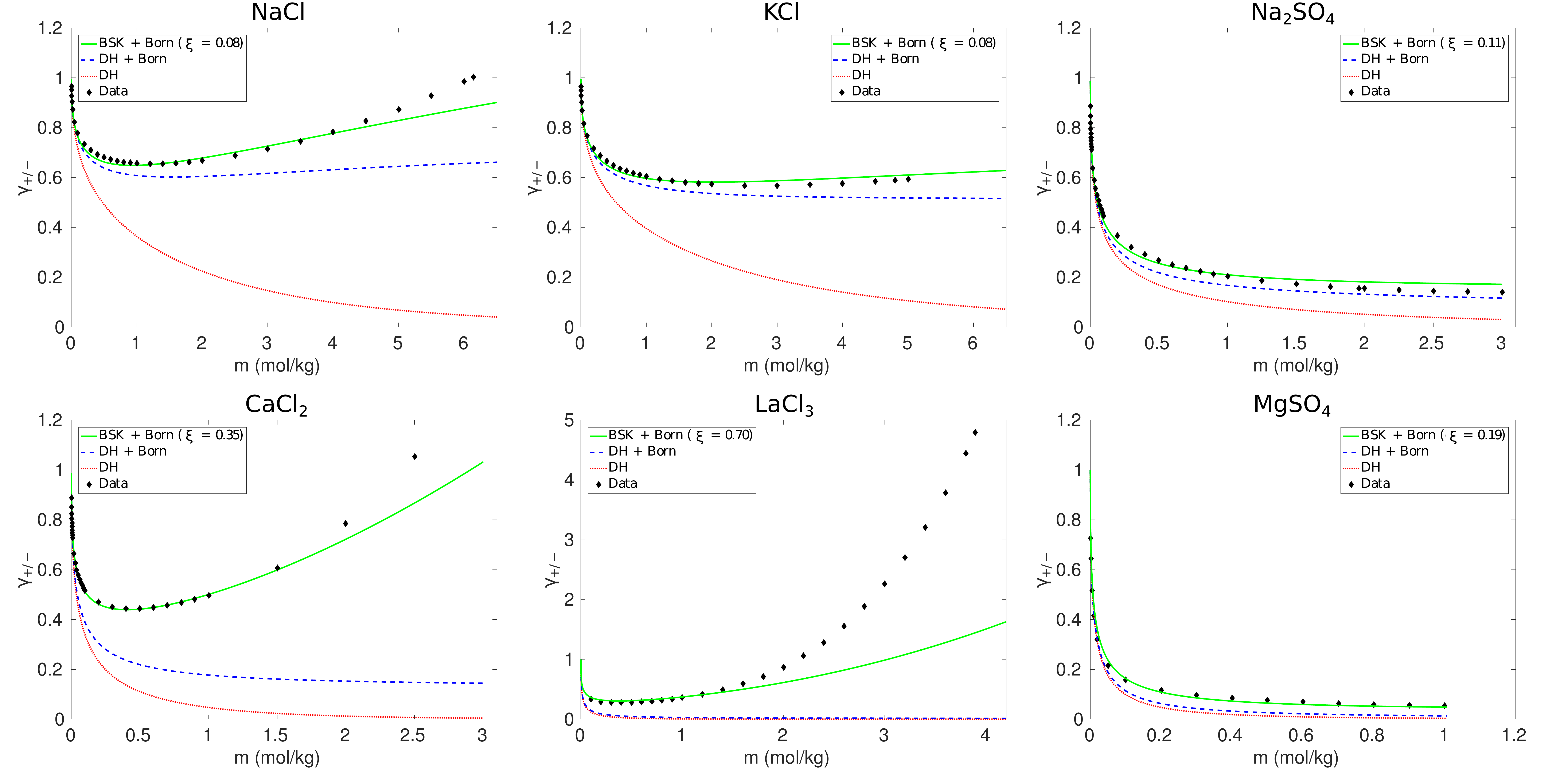}
\caption{ 
Experimental data~\cite{hamer_osmotic_1972,goldberg_evaluated_1981,staples_activity_1977,spedding_isopiestic_1976,bromley_thermodynamic_1973} for the mean activity coefficient, $\gamma_\pm = \left(\gamma_+^{s_+}\gamma_-^{s_-}\right)^{1/(s_++s_-)}$ versus molality at room temperature for six binary salts dissociating in water as $\mbox{A}_{s_+}\mbox{B}_{s_-} \rightleftharpoons s_+\mbox{A}^{z_+} + s_-\mbox{B}^{-z_-}$, compared with the predictions of Debye-H\"uckel theory (DH, Eq. ~(\ref{eq:DH})), H\"uckel's extension with Born solvation energy (DH $+$ Born, Eq. ~(\ref{eq:DHB})) and our theory (BSK $+$ Born, Eq. (\ref{eq:BSK})). Literature values were used for solution permittivities~\cite{zuber_empirical_2014} (Eq.~(\ref{eq:zuber}), densities \cite{haynes_concentrative_2012, novotny_densities_1988}, and ionic radii \cite{marcus_ionic_1988, volkov_two_1997}, $a=0.97$ (Na$^+$), 1.41 (K$^+$), 1.03 (Ca$^{2+}$), 0.70 (Mg$^{2+}$), 
1.80 (Cl$^-$), 2.42 (SO$_4^{2-}$) \AA. For each salt, our formula has only one adjustable parameter, $\xi$, to rescale the correlation length $\ell_c$ in Eq. (\ref{eq:lc}).  
\label{fig:goodfit}
}
\end{figure*}

After almost a century, the time is ripe to generalize DH for concentrated electrolytes, with the advent of an analytically tractable model for many-body Coulomb correlations.  Inspired by Santangelo's analysis of the one-component plasma at intermediate coupling~\cite{santangelo_computing_2006}, Bazant, Storey and Kornyshev (BSK) proposed a Ginzburg-Landau-like model of Coulomb correlations in ionic liquids~\cite{bazant_double_2011} and concentrated electrolytes~\cite{bazant_towards_2009,storey_effects_2012}, which leads to a fourth-order Poisson equation that predicts over-sceening phenomena in diverse situations~\cite{jiang2014dynamics,jiang2016current,stout2014continuum,alijo2014steric,kornyshev2014differential,yochelis2014spatial,elbourne2015nanostructure,kondrat2015dynamics,moon2015osmotic,wang2017modeling}.  Liu and Eisenberg solved the same equation numerically as two second-order equations~\cite{liu2015numerical} for nonlocal solvent polarization~\cite{xie2016nonlocal,kornyshev_nonlocal_1981, kornyshev_shape_1996, kornyshev_overscreening_1997} and applied it to ion channels~\cite{liu2013correlated} and ionic activity (including the Born energy)~\cite{liu_poisson_2015}, fitting activity data for NaCl and CaCl$_2$ with a correlation length close to the ion size.

In this Letter, we derive an activity formula based on BSK screening, augmented by the Born energy.  
We begin by comparing the classical theory with recent experiments. 
Zuber et al.~\cite{zuber_empirical_2014} fitted permittivity data to the empirical formula,
\begin{equation}
\varepsilon = \frac{\varepsilon_s}{1+\sum_i\alpha_ix_i} = \frac{\varepsilon_s}{1+\alpha x}   \label{eq:zuber}
\end{equation} 
where $\alpha_i$ is an ion-dependent parameter, $x_i$ is the mole fraction of ion $i$, $x$ is the mole fraction of the salt, and $\alpha = \sum_i s_i\alpha_i$.  Using measured permittivities, we show in  Fig. ~\ref{fig:goodfit} that adding the Born energy, Eq. (\ref{eq:DHB}), (without H\"uckel's linearization) greatly improves the fitting of activity data for concentrated solutions compared to the DH formula (\ref{eq:DH}). Interestingly, the largest discrepancies remain for multivalent cations (Ca$^{2+}$, Mg$^{2+}$, La$^{3+}$), which suggests the need to better account for Coulomb correlations.

The excess chemical potential can be approximated as a cluster expansion,
\begin{equation}
\mu_i^{ex} = \mu_i^0 + 2\pi \sum_{j} \int_{0}^{\infty}  K_{ij}(r) g_{ij}(r) \bar{c}_j r^2 dr + ...  \label{eq:integral}
\end{equation}
where we keep only the ion-solvent self energy, $\mu_i^0(\{ \bar{c}_j \})$, and the two-body energy, where $K_{ij}(r)$ is the ion-ion pair potential and $g_{ij}(r)$ is the pair correlation function.
Following H\"uckel, we set $\mu_i^0 = \Delta E^B_i$, and neglect entropic effects of ion crowding~\cite{bazant_towards_2009}, which become important at high concentrations~\cite{kilic2007a} and  in ionic  liquids~\cite{kornyshev2007}.  Following DH,  we consider only Coulomb forces, $K_{ij} = \frac{(z_ie)(z_je)}{4\pi\varepsilon r}$, neglect short-range (e.g. hydration) forces, and calculate $g_{ij} \approx 1 - \frac{z_ie\psi}{k_BT}$ for the screening cloud around central ion of radius $a_i$ in local equilibrium with a small fluctuating pair potential, $\psi(r)$.  

In place of the DH mean-field approximation, we capture some many-body correlations via the linearized BSK equation, which takes the dimensionless form,
\begin{equation}
\left(1-\delta_c^2\tilde{\nabla}^2\right)\tilde{\nabla}^2\tilde{\psi} = \tilde{\psi}
\label{eq:BSKeq}
\end{equation}
where $\tilde{\psi}=z_ie\psi/k_BT$ , $\tilde{\nabla}=\lambda_D \nabla$, and $\delta_c = \ell_c/\lambda_D$.  Motivated by BSK~\cite{bazant_double_2011}, we set the correlation length, $\ell_c$, proportional to the harmonic mean of the Bjerrum length, $\ell_B = \frac{e^2}{4\pi\varepsilon k_BT}$, in dilute solutions and the mean ion spacing, $c_0^{-1/3}$, in concentrated solutions,
\begin{equation}
\ell_c= 2\xi\left(\ell_B^{-1}+c_0^{1/3}\right)^{-1}    \label{eq:lc}
\end{equation}
with one adjustable parameter, $\xi$. The general decaying solution of (\ref{eq:BSKeq}) in spherical coordinates is
\begin{equation}
\tilde{\psi} = c_1\frac{e^{-b_1\tilde{r}}}{\tilde{r}}  + c_2\frac{e^{-b_2\tilde{r}}}{\tilde{r}} 
\label{psiE}
\end{equation}
where 
\begin{equation}
b_1 = \sqrt{\frac{1-\sqrt{1-4\delta_c^2}}{2\delta_c^2}}, \  b_2 = \sqrt{\frac{1+\sqrt{1-4\delta_c^2}}{2\delta_c^2}}.
\end{equation}
Gauss' law for nonlocal BSK polarization requires~\cite{bazant_double_2011},
\begin{equation}
-\hat{n}\cdot\left(1-\delta_c^2\tilde{\nabla}^2\right)\tilde{\nabla}\tilde{\psi}\left|_{\tilde{r}=\tilde{a_i}}\right.  = \tilde{q}_i
= \frac{z_ie^2\lambda_D}{4\pi a_i^2\varepsilon k_BT}
\label{BC}
\end{equation}
where $\tilde{q}_i$ is the dimensionless surface charge, $\tilde{a}_i=a_i/\lambda_D$, and $\hat{n}=\hat{r}$.  

The fourth-order Poisson equation requires another boundary condition, which controls electrostatic correlations at the surface. When this approach was first considered for electrolytes, Bazant et al.~\cite{bazant_towards_2009} proposed a mixed Stern-layer boundary condition, interpolating between 
\begin{equation}
\hat{n}\cdot\nabla\nabla^2\psi=0   \mbox{ (fixed potential)}  \label{eq:third}
\end{equation}
for a surface of fixed potential without adsorbed charge, as used in modeling electrode double layers~\cite{bazant_double_2011}, and 
\begin{equation}
\hat{n}\cdot\nabla\psi=0  \label{eq:first} \mbox{ (fixed charge)}
\end{equation}
for a surface of fixed charge, as derived by Santangelo~\cite{santangelo_computing_2006} for the one-component plasma near a charged wall. Since our spherical ion has fixed charge,  we choose Eq. (\ref{eq:first}) and obtain
\begin{equation}
c_1 = \frac{\tilde{q}_i\tilde{a}_i\exp{(b_1\tilde{a}_i)}}{\delta_c^2\left(b_1\tilde{a}_i+1\right)\left(b_2^2-b_1^2\right)}, \ 
c_2 = \frac{\tilde{q}_i\tilde{a}_i\exp{(b_2\tilde{a}_i)}}{\delta_c^2\left(b_2\tilde{a}_i+1\right)\left(b_1^2-b_2^2\right)}.
\end{equation}
Below we also consider the opposite limit of  Eq. (\ref{eq:third}). 

Using these results to perform the integral in Eq. (\ref{eq:integral}) and enforcing bulk electroneutrality, $\sum_i z_i s_i = 0$, we arrive at our main result:
\begin{equation}
\ln \gamma_i = \ln\gamma_i^{BSK} + \frac{\Delta E^B_i}{k_BT}, \ \ \mbox{where}  
\label{eq:BSK}
\end{equation}
\begin{equation}
\ln\gamma_i^{BSK} = \frac{z_i\tilde{q}_i\tilde{a}_i^2}{2\delta_c^2\left(b_1^2-b_2^2\right)}\left[\frac{1}{b_1(b_1\tilde{a}_i+1)}-\frac{1}{b_2(b_2\tilde{a}_i+1)}\right].    \nonumber
\end{equation}
As shown in Fig. ~\ref{fig:goodfit}, this simple formula is able to predict activity data for diverse aqueous electrolytes with remarkable accuracy, considering it has only one adjustable parameter, $\xi$. The fitted correlation lengths $\ell_c\propto\xi$ scale roughly with cation valence squared, $\xi \approx \xi_0 z_+^2$, which supports the BSK interpretation of Eq. (\ref{eq:BSKeq}) in terms of ion-ion correlations, as opposed to nonlocal solvent polarization with $\ell_c=$ constant~\cite{liu_poisson_2015}.  In hindsight, for aqueous solutions it is natural to multiply $\ell_c$ by $z_+^2$ in Eq. (\ref{eq:lc}) since cation correlations are favored by hydration chemistry, as negative oxygen atoms can order around cations more easily than do positive di-hydrogens around anions.  With this choice, we could fit all the data in Fig. ~\ref{fig:goodfit} with a single universal parameter, $\xi_0= 0.08$, which is comparable to the value suggested by BSK~\cite{bazant_double_2011,BSKerratum} for $\ell_c$ in the high concentration limit of ionic liquids, $\xi_0^{rcp} = \left(\frac{3\Phi^{rcp}}{4\pi}\right)^{1/3}=  \frac{1}{2(0.83)^{1/3}}=0.53$ for random close packing of spheres at volume fraction, $\Phi^{rcp}=0.63$.

\begin{figure}
\begin{center}
\includegraphics[keepaspectratio=true,width=3in]{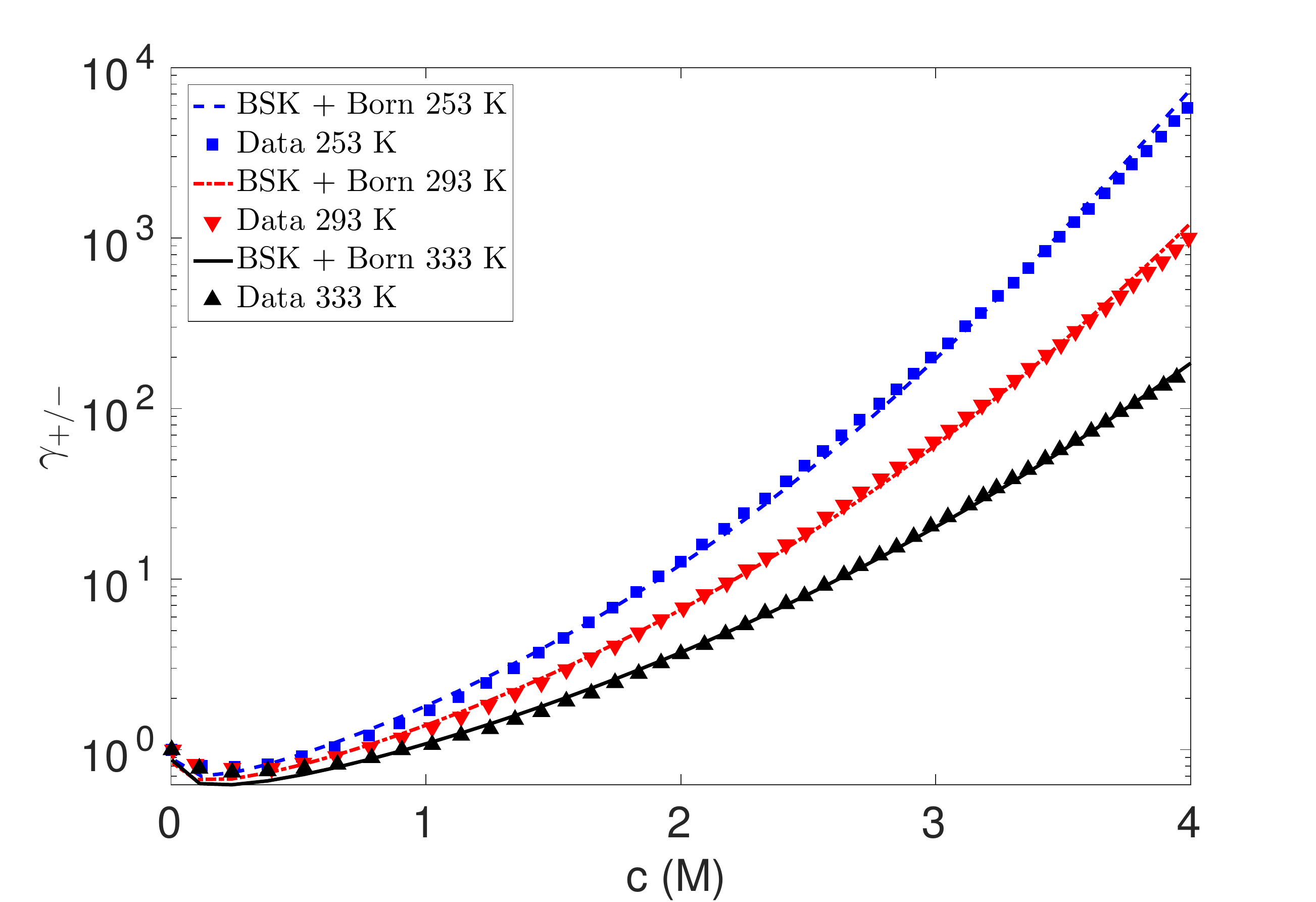}\\
	\begin{tabular}{ |c|c|c|c| } 
	\hline
	Temperature (K) & $\delta_s$ (m$^3$/mol) & b$_s$ (m$^{9/2}$/mol$^{3/2}$) & $\varepsilon_{solvent}$\\
	\hline
	253 & 1.830 $\times 10^{-2}$ & 1.606 $\times 10^{-4}$ & 40\\ 
	293 & 1.328 $\times 10^{-2}$ & 1.088 $\times 10^{-4}$ & 33\\ 
	333 & 9.863 $\times 10^{-3}$ & 7.595 $\times 10^{-5}$ & 28\\ 
	\hline
	\end{tabular}
\end{center}
\caption{ Predictions of Eq. (\ref{eq:BSK}) compared with data~\cite{valoen_transport_2005} (points) for a typical Li-ion battery electrolyte  (LiPF$_6$ in 10\% propylene, 27\% ethylene, 63\% dimethyl carbonates).  Since permittivity data is not available, we set $\xi=1$ and fit only the dielectric decrements, $\delta_c(T)$ and $b_s(T)$, to Eq. (\ref{eq:vinperm}) using known solvent permittivites~\cite{seward1958dielectric, chernyak2006dielectric,lee2013dielectric}, $\varepsilon_s(T)$.}
\label{fig:LiPF6}
\end{figure}

Our formula is also able to fit data for organic Li-ion battery electrolytes, as shown in Fig. ~\ref{fig:LiPF6}. Since activity coefficients were extracted from electrochemical signals by fitting an engineering model~\cite{valoen_transport_2005}, permittivities were not directly measured, so we reverse the procedure above by setting $\xi=1$ and fitting the data to Eq. (\ref{eq:BSK}) using the permittivity relation of Vincze et al.~\cite{vincze_nonmonotonic_2010},
\begin{equation}
\varepsilon(c_0,T) = \varepsilon_s(T) - \delta_s(T) c_0 + b_s(T) c_0^{3/2},  \label{eq:vinperm}
\end{equation} 
where we estimate solvent permittivity, $\varepsilon_s(T)$, from the literature values~\cite{seward1958dielectric, chernyak2006dielectric, lee2013dielectric}. With only two adjustable parameters, $\delta_s(T)$ and $b_s(T)$, our formula is able to fit the data across a wide range of temperatures and concentrations, and the inferred dielectric decrements, $\delta_s(T)$, are consistent with values for aqueous solutions~\cite{vincze_nonmonotonic_2010}.  

Besides providing a simple formula for fitting thermodynamic data, our theory also sheds light on the complex physics of screening.  At moderate dilution, activity increases from the DH dilute limit  (since $\gamma_i^{DH}<1$),
\begin{eqnarray}
\frac{\ln\gamma_i^{BSK}}{\ln\gamma_i^{DH}} 
&\sim& 1- \left(\frac{1+\tilde{a_i}+\tilde{a}_i^2}{\tilde{a}_i+\tilde{a}_i^2}\right) \delta_c^2 \   \mbox{ for } \delta_c\ll\frac{1}{2}
\end{eqnarray}
because the screening length, $\lambda_s = \lambda_D/(\mbox{Re }b_1)$, increases slightly  
$\tilde{\lambda}_s  \sim 1+ \delta_c^2 + \ldots$  for $\delta_c \ll 1$, thus reducing attractive Coulomb correlations.  In the opposite limit of high ionic strength, $I >(32\pi \ell_B \ell_c^2)^{-1}$ or $\delta_c > \frac{1}{2}$,  ``over-screening" charge oscillations arise~\cite{bazant_double_2011} with radial wavelength, $\lambda_o=\frac{2\pi \lambda_D}{\mbox{Im }b_1}$, and the screening length diverges, 
\begin{equation}
\tilde{\lambda}_s \sim \frac{\tilde{\lambda}_o}{2\pi} \sim \sqrt{2 \delta_c} \ \mbox { for }  \delta_c \gg \frac{1}{2},   \label{eq:lamhigh}
\end{equation}
yielding a critically damped long-range oscillation,  $\lambda_s \sim \frac{\lambda_o}{2\pi} \sim \sqrt{2 \ell_c \lambda_D}$. A similar phenomenon of ``underscreening"~\cite{lee2017underscreening,gebbie2017long} was recently discovered by surface-force measurements in ionic liquids~\cite{gebbie2013ionic,gebbie2015long} and concentrated electrolytes~\cite{smith2016electrostatic}  and attributed  to short-range attractive forces that trap mobile ions in Bjerrum pairs~\cite{gebbie2013ionic,adar2017bjerrum,goodwin2017underscreening,gavish2017solvent}, despite the high conductivity and double-layer capacitance~\cite{andersson2017nanotribology} and seemingly weak specific forces\cite{lee2014room}.  Interestingly, our theory predicts the observed universal scaling of the underscreening length, $\tilde{\lambda}_s$ vs $\tilde{a}_i \approx\delta_c$, albeit with smaller magnitude, $\lambda_s\approx1-10$nm, and exponent $\frac{1}{2}$  in Eq. (\ref{eq:lamhigh}) instead of $\approx 3$ (Fig. 4 of Smith et al. ~\cite{smith2016electrostatic}), based only on many-body Coulomb correlations among mobile ions.
Over-screening (charge oscillations) and underscreening (extended range) both weaken attractive Coulomb correlations, thereby increasing the activity relative to DH:
\begin{eqnarray}
\frac{\ln\gamma_i^{BSK}}{\ln\gamma_i^{DH}} 
&\sim& \frac{1+\tilde{a}_i}{\sqrt{2\delta_c}}\left(1 - \frac{1}{4\delta_c^2} \right) \  \mbox{ for } \delta_c\gg\frac{1}{2} \ \  
\end{eqnarray}
As shown in Fig. ~\ref{fig:goodfit}, however, the activity of concentrated electrolytes 
cannot be attributed solely to underscreening, as conjectured by Lee et al.~\cite{lee2017underscreening} and tested against data for NaCl without considering solvation energy or dielectric decrement.

\begin{figure}[t]
\begin{center}
\includegraphics[keepaspectratio=true,width=3in]{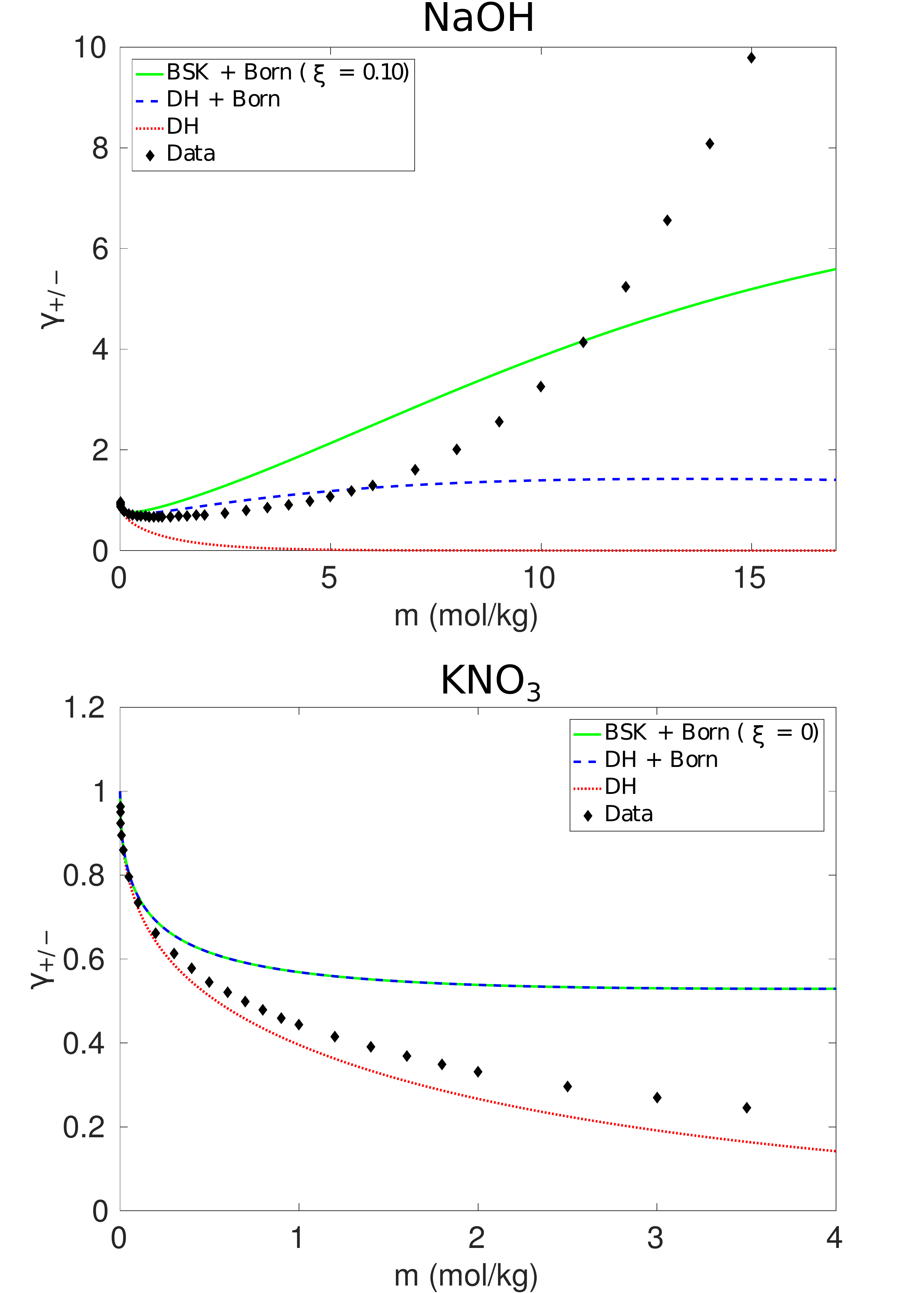}
\end{center}
\caption{ The same comparison of activity models with experimental data as in Fig. ~\ref{fig:goodfit} for two cases where our formula fails to capture strong short-range interactions involving certain ions, hydroxyl ($a=1.33$ \AA\ for OH$^-$) and nitrate ($a=1.77$ \AA\ for NO$_3^-$), at high concentrations.
}
\label{fig:badfit}
\end{figure}

Our analysis also supports the interpretation of boundary conditions for BSK theory in terms of ion-image correlations, as in colloids~\cite{hatlo2008role}.  Repeating the screening calculation with vanishing third-derivative (\ref{eq:third}), we obtain a modified formula~\cite{sven_thesis}, 
\begin{equation}
\ln\gamma_i^{mBSK} = \frac{z_i\tilde{q}_i\tilde{a}_i^2}{2\delta_c^2\left(b_1^2-b_2^2\right)}\left[\frac{b_2^2}{b_1(b_1\tilde{a}_i+1)}-\frac{b_1^2}{b_2(b_2\tilde{a}_i+1)}\right]   \label{eq:metal}
\end{equation}
which predicts lower activity than DH theory, the opposite trend of Eq. (\ref{eq:BSK}) for vanishing first derivative (\ref{eq:first}).  Equation (\ref{eq:first}) thus captures repulsive ion-image forces for an ideally non-polarizable surface of fixed charge, which raise the activity of an ion (\ref{eq:BSK}) compared to DH, while Equation (\ref{eq:third}) describes attractive ion-images forces for an ideally polarizable surface of fixed potential, which lower the activity. The modified BSK formula (\ref{eq:metal}) may find applications to metal nanoparticles, e.g. describing their solubility in ionic liquids~\cite{dupont2010structural}.  

Although our theory captures much of the physics of ionic activity, it neglects short-range specific interactions that become  important at high concentration (including the ``solvent-in-salt" limit~\cite{suo2013new,suo2015water}) especially for certain ions, as shown in Fig.~\ref{fig:badfit}.   For aqueous KNO$_3$, the DH$+$Born model over-estimates the mean activity coefficient, so the positive BSK correction for fixed-charge ions (\ref{eq:BSK}) cannot improve the fit.  Interestingly, the negative correction for an ideally polarizable``metallic ion" (\ref{eq:metal}) could fit the data, which seems consistent with the fact that nitrate ions have a labile hydration shell with hydrogen bonds fluctuating at the time scale as ion polarization~\cite{thogersen2013hydration}. In contrast, hydroxyl ions have a tightly bound hydration shell~\cite{botti2003solvation}, which could interfere with the Grotthuss mechanism of proton hopping~\cite{grothuss1806,bernal1933theory,agmon1995grotthuss} and raise the hydroxyl activity by lowering entropy,  as shown in Fig.~\ref{fig:badfit} for aqueous NaOH.  It is possible to extend our theory to include short-range interactions~\cite{levy2017prep} using the regular-solution approximation of Goodwin and Kornyshev~\cite{goodwin2017underscreening}, which makes the activity formula more cumbersome, but still physics-based, and thus more predictive than SIT or other empirical relations~\cite{may2017thermodynamic,zemaitis2010handbook,tester_thermodynamics_1997,bromley_thermodynamic_1973}. 

In conclusion, we arrive at a physical picture of ionic activity governed by three types of electrostatic forces: (i) mean-field ion-ion correlations (DH screening), (ii) ion-solvent self-energy (Born solvation), and (iii) many-body ion-ion and ion-image correlations (BSK screening).  Our activity formula (\ref{eq:BSK}) captures this physics in a simple way, which could be extended for nonlocal solvent polarization~\cite{levy2017prep,xie2016nonlocal,kornyshev_nonlocal_1981}, SIS discrete screening~\cite{fraenkel2010simplified,fraenkel2014computing}, solvent-mediated short-range forces~\cite{goodwin2017underscreening,levy2017prep}, and steric constraints~\cite{bazant_towards_2009}, with goal of describing the transition from concentrated electrolytes to  ionic liquids. 

This work was supported by a Professor Amar G. Bose Research Grant. The authors thank Amir Levy for references and helpful discussions.

\bibliography{ActivityPaper}



\end{document}